\documentclass[12pt,twoside,a4paper,useAMS]{meteoroids2013}
\usepackage{color,graphicx}
\usepackage[english]{babel}
\usepackage[round]{natbib}
\usepackage{url}
\usepackage{longtable}
\usepackage{lscape}
\title[New meteor showers identified]{New meteor showers identified in the CAMS and SonotaCo meteoroid orbit surveys}
\author[Rudawska \and Jenniskens]{Rudawska~R.$^1$ \and Jenniskens~P.$^2$}
\affiliation{$^1$Institut de M\'{e}canique C\'{e}leste et de Calcul des \'{E}ph\'{e}m\'{e}rides, 77 Avenue Denfert-Rochereau, 75014 Paris, France (email: rudawska@amu.edu.pl) \break
$^2$SETI Institute, 189 Bernardo Ave, Mountain View, CA 94043, USA
}
\jname{Proceedings of the Meteoroids 2013 Conference\\
       Aug. 26-30, 2013, A.M. University, Pozna\'{n}, Poland}
\editors{Jopek T.J., Rietmeijer F., Watanabe J., Williams I.P., ed.}
\begin{document}
\maketitle
\begin{abstract}
A cluster analysis was applied to the combined meteoroid orbit database derived from low-light level video observations by the SonotaCo consortium in Japan (64,650 meteors observed between 2007 and 2009) and by the Cameras for All-sky Meteor Surveillance (CAMS) project in California, during its first year of operation (40,744 meteors from Oct. 21, 2010 to Dec. 31, 2011). The objective was to identify known and potentially new meteoroid streams and identify their parent bodies. The database was examined by a single-linking algorithm using the Southworth and Hawkins D-criterion to identify similar orbits, with a low criterion threshold of D\,$<$\,0.05. A minimum member threshold of 6 produced a total of 88 meteoroid streams. 43 are established streams and 45 are newly identified streams. The newly identified streams were included as numbers 448-502 in the IAU Meteor Shower Working List. Potential parent bodies are proposed. 
\end{abstract}
\section{Introduction}
Ongoing meteoroid orbit surveys aim to identify as many as possible meteoroid streams. Each stream originated from a parent comet or asteroid. By integrating the parent body orbit back in time, dust can be generated at different epochs and then followed forward to its orbital evolution into an Earth intersecting stream at the present time to confirm the association. Each conclusive link provides a record of the parent body past activity and a 3-dimensional distribution of the dust in the inner solar system now and into the future ~(\cite{Jenniskens_2006}).

In recent years, networks of low-light-level video cameras have contributed many new meteoroid orbits, complimenting radar studies such as results from the Canadian Meteor Orbit Radar~(\cite{Brown_2010}). The most productive camera network has been that of the Japanese SonotaCo consortium (\cite{SonotaCo_2009}). Meteoroid stream searches in Europe have been mostly focused on single-station observations (e.g. the International Meteor Organization Video Meteor Database), but multi-station results are now being gathered by the European Video Meteor Network Database~(\cite{Kornos_2013}).

In California, the Cameras for All-sky Meteor Surveillance (CAMS) network started operations in October of 2010. At the end of 2011, 40,744 meteoroid orbits were calculated. The work presented here was our first attempt to confirm some of the previously reported showers listed in the IAU Working List of Meteor Showers, and to find potential new meteoroid streams. We did so by combining the CAMS data with those released by the SonotaCo consortium.

\section{Methods}
The 60-camera CAMS network consists of three stations of 20 low-light-level video cameras each, located at Fremont Peak Observatory, at Lick Observatory, and in Sunnyvale (or alternatively in Mountain View or Lodi) in California. The three stations used Watec Wat-902H2 Ultimate cameras equipped with f1.2 12~mm focal length lenses. The video is stored in four-frame compressed format and analysed using the CAMS software package~(\cite{Jenniskens_2011}). 

The 100-camera SonotaCo network in Japan consists of more than 30 stations, which use both WATEC-100N and WATEC-902H2 cameras equipped with 3.8~--~12~mm lenses (\cite{SonotaCo_2009}). Data are recorded and analyzed by the UFO software package. The precision of the measured radiant positions is about a factor of two less precise than the CAMS network (\cite{Jenniskens_2011}). At the time of this work, in early 2012, the 2007~--~2009 meteoroid orbits were made publicly available (64,650 meteors). 

\section{Meteor showers identification}
The combined database contains 105,394 meteoroid orbits. This meteor database was examined using a single-linking orbit grouping method by means of the Southworth and Hawkins criterion (\cite{D_SH}) to identify similar orbits. 

In a first step, the IAU List of Established Meteor Showers was used to identify known meteor showers in the combined database. For each stream identified in this step the mean orbit was calculated and stream members identified. These orbits were excluded from the subsequent group-search using the single-linking method. Our grouping algorithm (\cite{Rudawska_2012}) is based on the single linkage, or nearest neighbour, method proposed by Southworth and Hawkins (1963). In addition, they introduced the dissimilarity function, $D_{SH}$, which now is the most often used criterion to identify what orbits are similar and may be linked.

Two orbits are thought associated if D is less than an assumed constant threshold, often taken as D\,$<$\,0.25 (\cite{Lindblad_1971}). In the first step, in which the major showers were identified, we used a threshold of 0.10. In the subsequent grouping algorithm, we used a low value of 0.05. Hence, only the most identical orbit groupings were extracted from the surveys. For our preliminary parent body search, we used a higher threshold of 0.20. 

\section{Results}
\begin{figure}[]
\centering
\includegraphics[width=0.49\textwidth, trim= 13mm 66mm 15mm 91mm, clip]{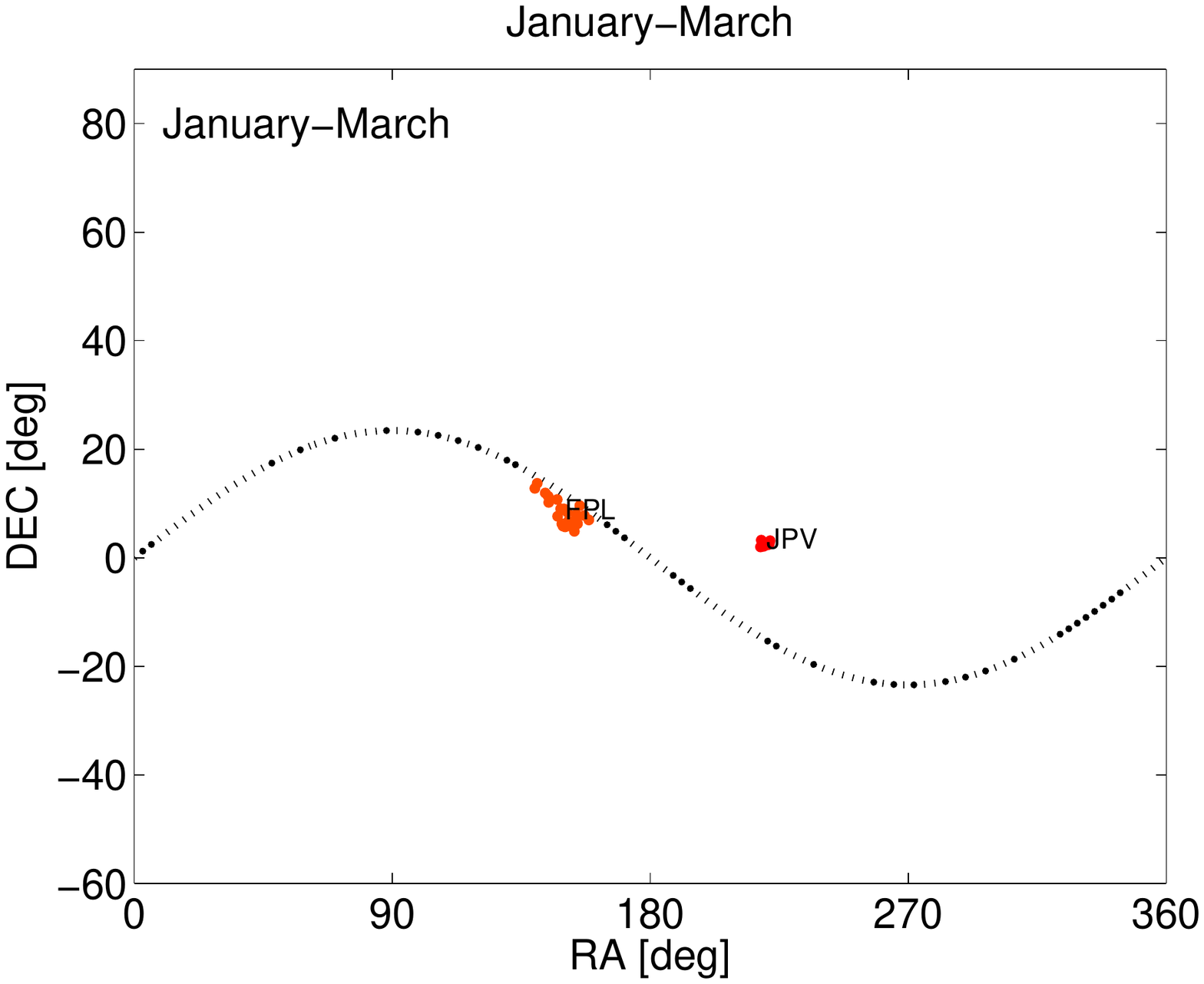}
\includegraphics[width=0.49\textwidth, trim= 13mm 66mm 15mm 91mm, clip]{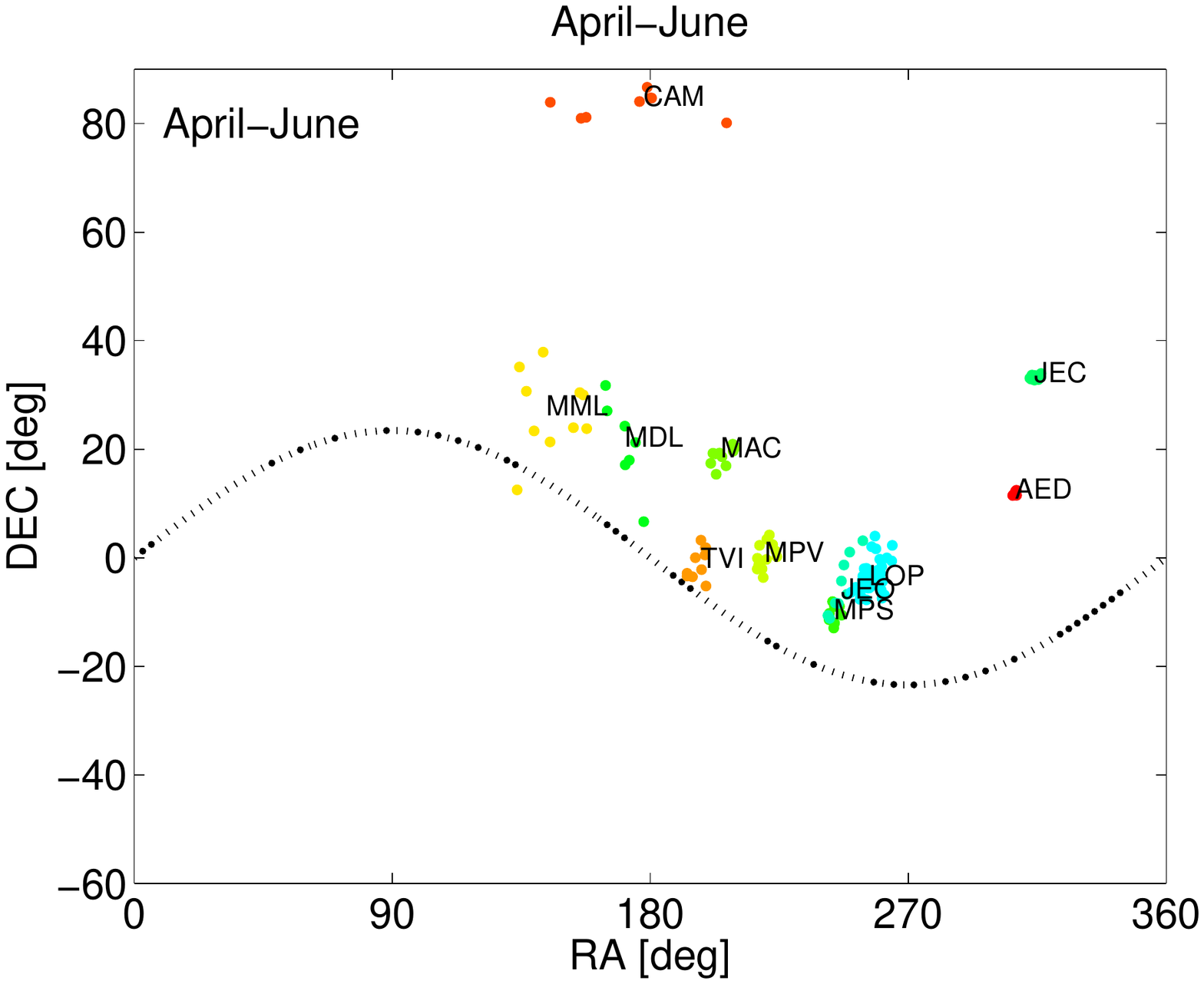}
\footnotesize
\caption{The radiant position of meteor showers found between January and June. The dashed line is the ecliptic plane.}
\label{fig:January-June}
\normalsize
\end{figure}
\begin{figure}[]
\centering
\includegraphics[width=0.49\textwidth, trim= 13mm 66mm 15mm 91mm, clip]{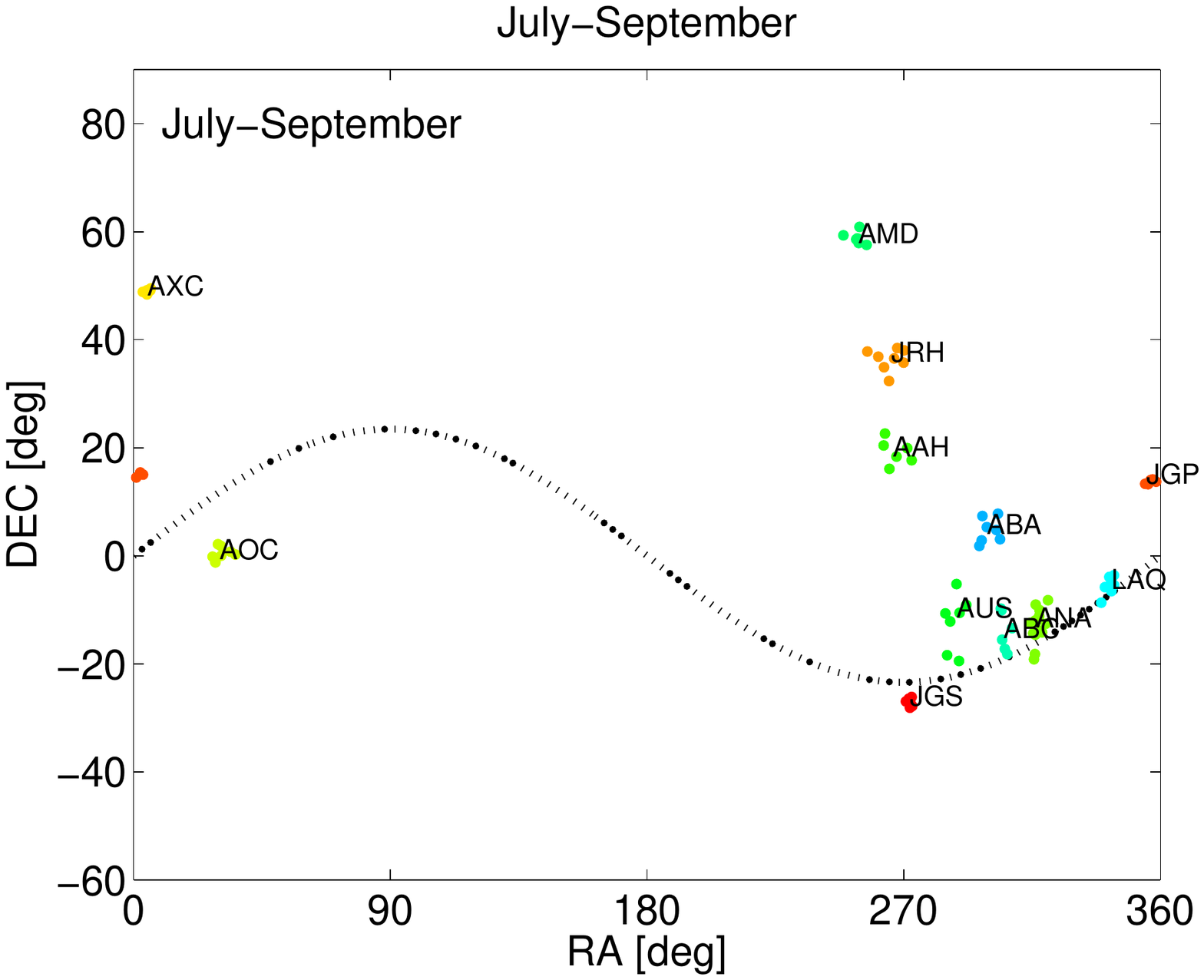}
\includegraphics[width=0.49\textwidth, trim= 13mm 66mm 15mm 91mm, clip]{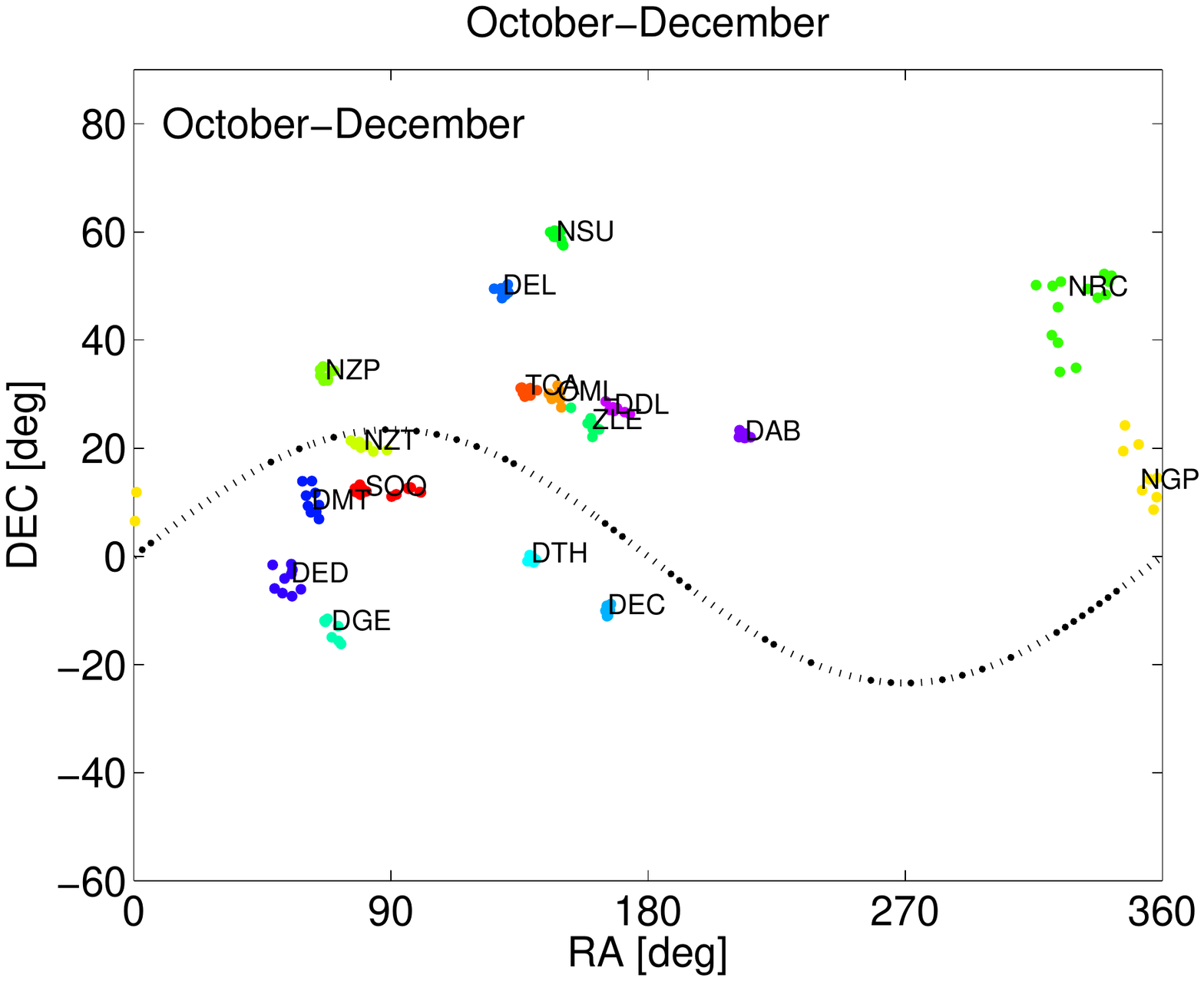}
\footnotesize
\caption{The radiant position of meteor showers found between July and December. The dashed line represents ecliptic plane.}
\label{fig:July-December}
\normalsize
\end{figure}
\begin{table}
 \caption{Mean orbital elements and radiant position of the newly identified meteoroid streams.}
 \label{tab:meanOrbRad}
 \centering
 \scriptsize
 \begin{tabular}{rr|rrrrr|rrrr|r|r}
  \hline
   IAU & Code & $e$ & $q$ & $i$ & $\omega$ & $\Omega$ & $\lambda_\odot$ & $\alpha$ & $\delta$ & $V_g$ & No & Name \\
  \hline
*448 & ALL &      0.969 &    0.070 &    9.2 &  332.8 &   13.6 &  13.6 &  219.7 &  -12.9 &   40.7 &          6 & April $\alpha$ Librids \\
 449 & ABS &      0.605 &    0.899 &    1.5 &   16.1 &  187.5 &  13.6 &  164.9 &   +4.4 &   12.3 &          7 & April $\beta$ Sextantids \\
*450 & AED &      0.997 &    0.745 &  122.3 &  119.1 &   20.1 &  20.2 &  307.2 &  +11.8 &   61.5 &          6 & April $\epsilon$ Delphinids\\
*451 & CAM &      0.619 &    0.998 &   20.7 &  168.1 &   39.1 &  39.1 &  172.6 &  +83.7 &   14.7 &          8 & Camelopardalids \\
*452 & TVI &      0.683 &    0.856 &    2.4 &  230.5 &   39.6 &  39.5 &  196.9 &   -1.0 &   14.9 &         10 &  $\theta$ Virginids \\
*453 & MML &      0.545 &    1.004 &    2.8 &  183.0 &   31.1 &  40.3 &  145.5 &  +27.2 &    7.7 &         10 & May  $\mu$ Leonids \\
*454 & MPV &      0.744 &    0.652 &   10.4 &  259.7 &   41.6 &  41.6 &  220.2 &   +0.3 &   21.7 &         12 & May  $\phi$ Virginids \\     
*455 & MAC &      0.612 &    0.902 &   10.0 &  223.4 &   42.8 &  42.8 &  205.2 &  +18.7 &   13.7 &         10 & May  $\alpha$ Comae Berenicids \\
*456 & MPS &      0.805 &    0.522 &    9.4 &  275.5 &   61.5 &  61.5 &  244.5 &  -10.6 &   25.4 &         10 & May  $\psi$ Scorpiids \\
*457 & MDL &      0.488 &    1.003 &    3.1 &  189.1 &   54.2 &  54.1 &  171.2 &  +20.9 &    7.4 &          7 & May  $\delta$ Leonids \\
*458 & JEC &      0.997 &    0.914 &   95.6 &  216.8 &   82.3 &  82.3 &  314.3 &  +33.2 &   53.2 &          9 & June  $\epsilon$ Cygnids \\ 
*459 & JEO &      0.656 &    0.833 &    7.3 &  236.8 &   84.1 &  84.1 &  246.9 &   -5.2 &   15.6 &          9 & June  $\epsilon$ Ophiuchids \\ 
 460 & LOP &      0.742 &    0.741 &   11.1 &  248.2 &   84.5 &  84.5 &  257.2 &   -3.7 &   19.7 &         37 &  $\lambda$ Ophiuchids \\ 
 461 & JGS &      0.692 &    0.796 &    1.8 &   61.8 &  288.8 & 108.9 &  272.1 &  -27.1 &   16.3 &          7 & July  $\delta$ Sagittariids \\
462=175 & JGP &      0.904 &    0.524 &  150.0 &  273.8 &  120.8 & 120.8 &  358.6 &   +14.2 &   62.6 &      8 & July  $\delta$ Pegasids \\ 
*463 & JRH &      0.633 &    0.981 &   21.3 &  203.8 &  124.6 & 124.6 &  265.1 &  +36.4 &   15.6 &          8 & July  $\rho$ Herculids \\ 
 464 & KLY &      0.698 &    0.939 &   24.7 &  215.1 &  126.8 & 126.9 &  277.5 &  +33.3 &   18.6 &          6 & $\kappa$ Lyrids \\
*465 & AXC &      0.887 &    0.907 &  104.9 &  219.1 &  135.8 & 135.8 &    4.9 &  +48.9 &   55.5 &          9 & August  $\xi$ Cassiopeiids \\ 
 466 & AOC &      1.016 &    0.724 &  157.9 &   64.2 &  318.5 & 138.5 &   30.7 &   +0.3 &   67.5 &          8 & August $o$ Cetids \\ 
*467 & ANA &      0.781 &    0.618 &    2.6 &  223.3 &  154.0 & 139.4 &  317.1 &  -13.1 &   21.8 &         13 & August  $\nu$ Aquariids \\
 468 & AAH &      0.598 &    0.987 &   11.9 &  200.6 &  142.1 & 142.1 &  267.2 &  +19.2 &   11.2 &          6 & August  $\alpha$ Herculids \\
 469 & AUS &      0.601 &    0.930 &    3.0 &  217.8 &  145.1 & 145.1 &  287.9 &  -12.2 &   11.1 &          7 & August  $\upsilon$ Sagittariids \\
*470 & AMD &      0.654 &    1.011 &   30.3 &  177.2 &  145.4 & 145.4 &  253.7 &   +58.8 &   19.5 &         6 & August  $\mu$ Draconids \\ 
 471 & ABC &      0.638 &    0.846 &    2.1 &  234.5 &  146.5 & 146.4 &  305.4 &  -14.1 &   13.9 &          6 & August  $\beta$ Capricornids \\
 472 & ATA &      0.648 &    0.790 &    7.4 &  243.5 &  147.3 & 147.3 &  310.6 &   -1.8 &   15.9 &          7 & August $\theta$ Aquilids \\
 473 & LAQ &      0.877 &    0.297 &    2.6 &  300.3 &  148.0 & 147.6 &  342.3 &   -5.5 &   30.6 &          9 &  $\lambda$ Aquariids \\
 474 & ABA &      0.701 &    0.872 &   10.2 &  228.1 &  148.7 & 148.7 &  300.0 &   +4.7 &   15.1 &          7 & August  $\beta$ Aquariids \\
 475 & SAQ &      0.709 &    0.723 &    1.3 &  215.1 &  165.6 & 158.1 &  329.2 &  -11.0 &   17.9 &         10 & September Aquariids \\
 476 & ICE &      0.832 &    0.421 &    3.4 &  134.0 &  357.2 & 176.1 &    4.8 &   -1.4 &   26.9 &          8 & $\iota$ Cetids \\
*477 & SRP &      0.723 &    0.726 &    6.1 &  249.8 &  177.4 & 177.4 &  343.4 &   +5.0 &   18.3 &          7 & September $\rho$ Pegasids \\
 478 & STC &      0.606 &    0.938 &    1.7 &  185.7 &  174.6 & 177.7 &  316.0 &  -14.1 &   10.3 &         13 & September $\theta$ Capricornids \\
*479 & SOO &      0.928 &    0.774 &  159.3 &   58.1 &    5.6 & 185.6 &   79.2 &  +12.1 &   67.6 &         18 & September $o$ Orionids \\ 
*480 & TCA &      0.822 &    0.845 &  154.8 &  131.8 &  207.3 & 207.3 &  137.5 &  +30.5 &   67.2 &          8 &  $\tau$ Cancrids \\ 
*481 & OML &      0.819 &    0.889 &  151.5 &  140.3 &  218.6 & 218.6 &  148.0 &  +29.7 &   67.3 &          7 & October  $\mu$ Leonids \\ 
*482 & NGP &      0.705 &    0.938 &    4.4 &  207.7 &  228.4 & 228.4 &  354.9 &  +14.4 &   11.4 &         10 & November $\gamma$ Pegasids \\
 483 & NAS &      1.103 &    0.898 &  154.2 &  325.5 &   51.4 & 231.4 &  149.9 &   -3.4 &   71.1 &          8 & November  $\alpha$ Sextantids \\ 
 484 & IOA &      0.652 &    0.836 &    3.4 &  193.7 &  200.4 & 234.7 &   27.6 &  +17.3 &   13.8 &        176 & $\iota$ Arietids \\
*485 & NZT &      0.926 &    0.172 &    5.5 &  135.9 &   60.0 & 240.0 &   81.0 &  +20.5 &   35.5 &          8 & November  $\zeta$ Taurids \\
*486 & NZP &      0.859 &    0.389 &   13.3 &  288.2 &  240.4 & 240.4 &   67.3 &  +33.6 &   29.4 &          9 & November  $\zeta$ Perseids \\ 
 487 & NRC &      0.616 &    0.982 &   14.4 &  189.1 &  241.4 & 241.4 &  323.2 &  +43.4 &   11.8 &         14 & November  $\rho$ Cygnids \\                                                                              
*488 & NSU &      0.972 &    0.813 &   99.9 &  230.0 &  241.6 & 241.6 &  148.3 &  +59.2 &   55.3 &         10 & November  $\sigma$ Ursae Majorids \\ 
 489 & ZLE &      1.344 &    0.953 &  155.3 &  200.0 &  248.6 & 248.6 &  159.6 &  +24.4 &   74.0 &          7 &  $\zeta$ Leonids \\ 
*490 & DGE &      0.818 &    0.716 &   23.5 &   66.8 &   69.3 & 249.3 &   69.5 &  -13.6 &   23.8 &          7 & December  $\delta$ Eridanids \\ 
 491 & DCC &      0.942 &    0.387 &  167.8 &  104.9 &   69.5 & 249.5 &  131.7 &  +12.8 &   64.1 &          6 & December  $\delta$ Cancrids \\
 492 & DTH &      0.967 &    0.695 &  149.6 &   66.3 &   71.9 & 251.9 &  139.2 &   -0.4 &   67.1 &          6 & December  $\theta$ Hydrids \\ 
 493 & DEC &      1.016 &    0.925 &  154.3 &  331.4 &   71.9 & 251.9 &  166.0 &  -10.0 &   70.6 &          6 & December  $\epsilon$ Craterids \\ 
*494 & DEL &      0.947 &    0.365 &   94.0 &  287.2 &  253.1 & 253.1 &  129.2 &  +49.1 &   51.7 &          8 & December Lyncids \\ 
 495 & DMT &      0.609 &    0.831 &    4.1 &   53.6 &   82.1 & 262.1 &   62.4 &  +10.3 &   13.5 &          9 & December $\mu$ Taurids \\
 496 & DED &      0.633 &    0.903 &    7.4 &   38.0 &   82.6 & 262.6 &   53.6 &   -4.4 &   12.2 &          9 & December  $\epsilon$ Eridanids \\
*497 & DAB &      1.002 &    0.686 &  112.3 &  113.3 &  263.9 & 263.9 &  213.5 &  +22.3 &   59.5 &          7 & December  $\alpha$ Bootids \\ 
*498 & DNH &      0.983 &    0.930 &  123.5 &   27.2 &   84.8 & 264.8 &  152.7 &  -23.8 &   63.8 &          7 & December  $\mu$ Hydrids \\
499=20 & DDL &      1.152 &    0.611 &  137.3 &  253.0 &  275.9 & 275.9 &  168.8 &   +27.2 &   67.0 &       8 & December  $\delta$ Leonids \\ 
*500 & JPV &      0.950 &    0.669 &  145.3 &  110.3 &  285.6 & 285.6 &  220.3 &   +2.5 &   66.2 &          7 & January  $\phi$ Virginids \\ 
 501 & FPL &      0.829 &    0.388 &    4.1 &  110.0 &  137.3 & 317.3 &  147.8 &   +9.0 &   28.3 &         14 & February  $\pi$ Leonids \\
*502 & DRV &      0.923 &    0.777 &  153.9 &  124.0 &  252.5 & 252.5 &  185.1 &  +12.9 &   68.1 &         19 & December $\rho$ Virginids \\
  \hline
 \end{tabular}
 \normalsize
\end{table}

The results for all clusters with at least 6 members (again, for threshold D\,$<$\,0.05), excluding previously identified meteoroid streams, are summarized in Table~\ref{tab:meanOrbRad}. The table shows the mean orbital elements and the mean radiant of each new meteoroid stream. The first two columns give the assigned IAU number and code of the stream. The next five columns show the mean orbital elements: eccentricity ($e$), perihelion distance ($q$), inclination ($i$), argument of perihelion ($\omega$), and longitude of ascending node ($\Omega$), respectively. The following four columns include the solar longitude, the right ascension and declination of the radiant, and the geocentric velocity. The twelfth column shows the number of identified meteors in each stream. The final column gives the name of the shower.

In the period between April and June we found 11 new meteoroid streams (Figure~\ref{fig:January-June}). 
In the period between April 22 and May 6, we identified the Camelopardalids shower (\#451), which may originate from comet 209P/LINEAR (formerly known as 2004~CB). This comet has a close encounter with Earth during May 2014, when Earth will cross potential dust trails from past returns. The shower provides some evidence that this weakly active comet produced large dust grains in the past. The predicted radiant of the possible outburst meteors in 2014 is at R.A.=125, Decl.=+78, and entry speed is 15.9~km/s (\cite{Jenniskens_2006}). The observed radiant is at R.A.=172.6, Decl.=+83.7 with entry speed of 14.7~km/s (Table~\ref{tab:meanOrbRad}).

From April to June, we found 11 new meteoroid streams (Figure~\ref{fig:January-June}). Meteoroid orbits of the April $\epsilon$ Delphinids (\#450) and the June $\epsilon$ Cygnids (\#458) are retrograde with unknown long-period comet parent bodies. These are compact showers.

The low velocity $\theta$ Virginids (\#452) may originate from 2011 HP4. The May $\mu$ Leonids (\#453) may derive from either 2009~EF21 or 2008~EH. The May $\phi$ Virginids may originate from 2005~JU1. The May $\psi$ Scorpiids (\#456) are a good match to the present orbit of 2009~KM. The May $\delta$ Leonids may originate from asteroid 2013~KB.

The May\,$\alpha$\,Comae\,Berenicids (\#455) move in an orbit roughly aligned with that of comet 73P/Schwassmann-Wachmann 3. The showers \#459 (June $\epsilon$ Ophiuchids) and \#460 ($\lambda$ Ophiuchids) may be the same shower, perhaps from Jupiter-family comet P/2005~JQ5 (Catalina), with $D_{SH}\,\approx\,$0.06. Shower 459 is the more certain one.

Between July and September we found 12 meteor showers. Compared to the spring meteor showers, these showers have a smaller number of members. With one exception, the August $\nu$ Aquariids (\#467), which includes 13 meteoroid orbits that cover a period between August 8 and 16. These may originate from comet 72P/Denning-Fujikawa. 

The July $\rho$ Herculids (\#463) are perhaps from asteroid 2011~MC. The retrograde August $\xi$ Cassiopeiids (\#465) may originate from a Halley-type comet. There is no candidate parent body. The September $\rho$ Pegasids (\#477)could derived from either asteroids 2011~EU29, 2004~NL8, or 2009~DA1. The September $o$ Orionids may originate from comet P/2005 T4 (SWAN).

The autumn showers (Figure~\ref{fig:July-December}) extractions include the $\iota$ Aquariids (\#484), with 176 meteors an unusually large group and possibly composed of multiple streams. Others are more typical. A few are highly inclined streams with retrograde orbits such as the December Lyncids (\#494). None have candidate parent bodies. The November $\gamma$ Pegasids (\#482) are possibly related to asteroids 2010~UM7, 2012~UU169, or 2007~PA8, all potentially dormant Jupiter Family comets. The Tisserand parameter for 2007~PA8 is 2.95 and for 2010~UM7 is 2.94, respectively. Moreover, the 2007 PA8 diameter is over 5 km in diameter. It would be interesting to study this object in reference to its possible past activity that may have supplied this meteoroid stream.

In January, we found the January $\phi$ Virginids (\#500), which is represented mostly by SonotaCo meteor orbits due to lack of CAMS observations at this time. The shower is active from January 3 till 7, peaking at solar longitude 285.6 degrees.
\vspace*{-.5cm}

\section{Discussion}
Because the CAMS survey is ongoing, these are preliminary results. In preparation of publication, these detections were reported to the IAU Meteor Data Center. To be more certain that these single-linking D-criterion extractions represented true meteoroid streams, we aimed to collect sufficient data within the CAMS network alone to confirm the detections. By the end of 2012, a total of 101,000 meteoroid orbits were measured by CAMS alone and the streams were again investigated, case by case. This work is the topic of a future publication. 

In total, 29 out of the 48 candidate showers in Table~\ref{tab:meanOrbRad} were confirmed. These are marked with a star in the first column of Table~\ref{tab:meanOrbRad}. In addition, we found that two of the newly identified streams are likely part of now established showers: Shower 462 = 175 and shower 499 is part of 20.

A total of 21 showers were not detected in the CAMS data collected so far (those without a star in Table~\ref{tab:meanOrbRad}). Based in part on SonotaCo data, it is possible that some of these showers were active only at some time in the period 2007-2009, before CAMS operations were started, or that they peak at solar longitudes not yet covered in the current CAMS observations due to bad weather on those days.

On the other hand, it is possible that, in spite of the low D threshold value used in this study, the remainder are mere chance associations of unrelated orbits. Single-linking of related orbits alone does not guarantee association. 

More meteoroid streams may exist in the data. The discriminant criterion, used at this low threshold value, does not select streams dispersed significantly in longitude of perihelion, for example. 

\section{Conclusions}
The Cameras for All-sky Meteor Surveillance (CAMS) and SonotaCo Network Japan meteor databases were examined by single-linking algorithm combined with Southworth and Hawkins D-criterion. A total of 88 meteor meteoroid streams were found, 43 of which were known established showers, two others have since been established. A significant fraction of the remainder (at least 29 out of 46) are newly recognized meteoroid streams. Potential parent bodies are proposed. 

\begin{acknowledgments} We thank Quentin Nenon (SETI Institute) for support of the stream verifications from the more recent CAMS data, and Jeremie Vaubaillon (IMCCE) for providing a postdoctoral position to RR and for facilitating a working visit by PJ to IMCCE. CAMS is supported by the NASA Near Earth Object Observation program.
\end{acknowledgments}
\bibliographystyle{meteoroids2013}
\bibliography{paperNewShowers}
%%%%%%%%%%%%%%%%%%%%%%%%%%%%%%%%%%%%%%%%%%%%%%%%%%%%%%%%%%%%%%%%%%%%%%%%%%%%%%
%
\end{document}